\documentclass[11pt,twocolumn,superscriptaddress]{revtex4-1}   
\pdfoutput=1
\usepackage{amsmath}
\usepackage{graphicx}
\usepackage{natbib}
\usepackage{mathrsfs}
\usepackage{ulem}
\usepackage{hyperref}
\usepackage[T1]{fontenc}
\usepackage[letterpaper,textwidth=7in,top=.75in,bottom=.75in]{geometry}
\usepackage{color}
\usepackage[dvipsnames]{xcolor}

\usepackage{soul}

\begin{document}
\title{Effect of inter-system crossing rates and optical illumination on the polarization of nuclear spins nearby nitrogen-vacancy centers}


 \author{H. Duarte}
 \address{Facultad de F\'isica, Pontificia Universidad Cat\'olica de Chile, Santiago 7820436, Chile}
    \address{Faculty of Physics, Universidad T\'ecnica Federico Santa Mar\'ia, Avda.~Vicu\~{n}a Mackenna 3939, Santiago, Chile}

 \author{H. T. Dinani}
  \address{Facultad de F\'isica, Pontificia Universidad Cat\'olica de Chile, Santiago 7820436, Chile}
      \address{
     Centro de Investigaci\'{o}n DAiTA Lab, Facultad de Estudios Interdisciplinarios,\\ Universidad Mayor, 	Santiago, Chile}

 \author{V. Jacques}
    \address{
     L2C, Laboratoire Charles Coulomb, Universit\'{e} de Montpellier and CNRS, Montpellier 34095, France}

 \author{J. R. Maze}
  \address{Facultad de F\'isica, Pontificia Universidad Cat\'olica de Chile, Santiago 7820436, Chile}
    \address{Research Centre for Nanotechnology and Advanced Materials, Pontificia Universidad Cat\'olica de Chile, Santiago, Chile}

\date{\today}

\begin{abstract}
Several efforts have been made to polarize the nearby nuclear environment of nitrogen-vacancy (NV) centers for quantum metrology and quantum information applications. Different methods showed different nuclear spin polarization efficiencies and rely on electronic spin polarization associated to the NV center, which in turn crucially depends on the inter-system crossing. Recently, the rates involved in the inter-system crossing have been measured leading to different transition rate models. Here, we consider the effect of these rates on several nuclear polarization methods based on the level anti-crossing, and precession of the nuclear population while the electronic spin is in the $m_s=0$ and $m_s=1$ spin states. We show that the nuclear polarization depends on the power of optical excitation used to polarize the electronic spin. The degree of nuclear spin polarization is different for each transition rate model. Therefore, the results presented here are relevant for validating these models and for polarizing nuclear spins. Furthermore, we analyze the performance of each method by considering the nuclear position relative to the symmetry axis of the NV center. 
\end{abstract}

\keywords{nuclear spin polarization, diamond, nitrogen-vacancy center, quantum information, solid-state nuclear polarization}

\maketitle

\section{Introduction}
Nuclear spins are promising candidates for storing and processing quantum information \cite{Fuchs, Maurer} due to their isolation and consequential long coherence times. For such applications, the state of nuclear spins must be initialized and read out with high fidelity, a difficult task due to the small magnetic moment of nuclear spins. However, they can be accessed through an ancillary electronic spin. For instance, carbon ($^{13}$C with spin 1/2) and nitrogen ($^{14}$N with spin 1 or $^{15}$N with spin $1/2$) in diamond are accessible through the ancillary electron spin of the nitrogen-vacancy (NV) center \cite{Neumann, Raularxiv} by optical means and through several methods. The performance of these methods crucially depend on the dynamics of the NV electronic spin. 

The electronic spin of NV centers have been widely used in quantum metrology and quantum information processing \cite{Maze, Rondin, Bonato, HTD, HTDnanomat, Wrachtrup} due to its long coherence time in a wide range of temperatures, and its accessibility through optical excitation \cite{Bala, Chu}. It is clear that its optical readout is the result of a spin-dependent inter-system crossing involving metastable singlet states \cite{Doherty_physrep}. However, although several models have been proposed to describe the optical excitation of the center and transitions to and from the singlet states \cite{Manson, Robledo, Tetienne, Gupta, Hacquebard}, it is still not clear which model is valid.

As it will be discussed here, the electron spin polarization can be transferred to nearby nuclear spins \cite{Dutt, Vincent, Fischer, Busaite, Bajaj, Jamonneau_thesis, Schwartz}. This has been used to hyperpolarize diamond particles \cite{Ajoy}, and even external species on the diamond surface using shallow implanted NV centers \cite{Fernandez}. Here, we focus on the effect of the inter-system crossing for transferring the electron spin polarization to nearby nuclear spins based on three methods: the excited state level anti-crossing (ESLAC) \cite{Vincent}, precession of nuclear spin while the electron spin is in its $m_s=0$ \cite{Dutt}, and while in $m_s=1$ spin projections \cite{Jamonneau_thesis, YunNJP, ZhangPhysRevAppl}. Although several other methods exists to polarize nuclear spins, we focus only on these three methods to illustrate the effect of the inter-system crossing rates.

In particular, we investigate how four different transition rate models \cite{Manson, Robledo, Tetienne, Gupta} impact the polarization of nuclear spins nearby the NV center for these three methods. Moreover, we compare the performance of these three methods by considering the nuclear position relative to the symmetry axis of the NV center as several experimental studies have observed different polarization for different nuclei. We first analyze the effect of different transition rate models on the degree of polarization of the electronic spin. Then, Section \ref{secNuclear} describes three different methods for achieving nuclear spin polarization. We give special attention to the role of the electronic spin on the nuclear polarization dynamics for each method. Finally, in Section \ref{secTheta} we discuss the nuclear spin polarization efficiency under different polar position of the nuclear spin relative to the NV symmetry axis.

\begin{figure}[t]
\hspace{-0.5cm}
\centering
    \includegraphics[width=.49\textwidth]{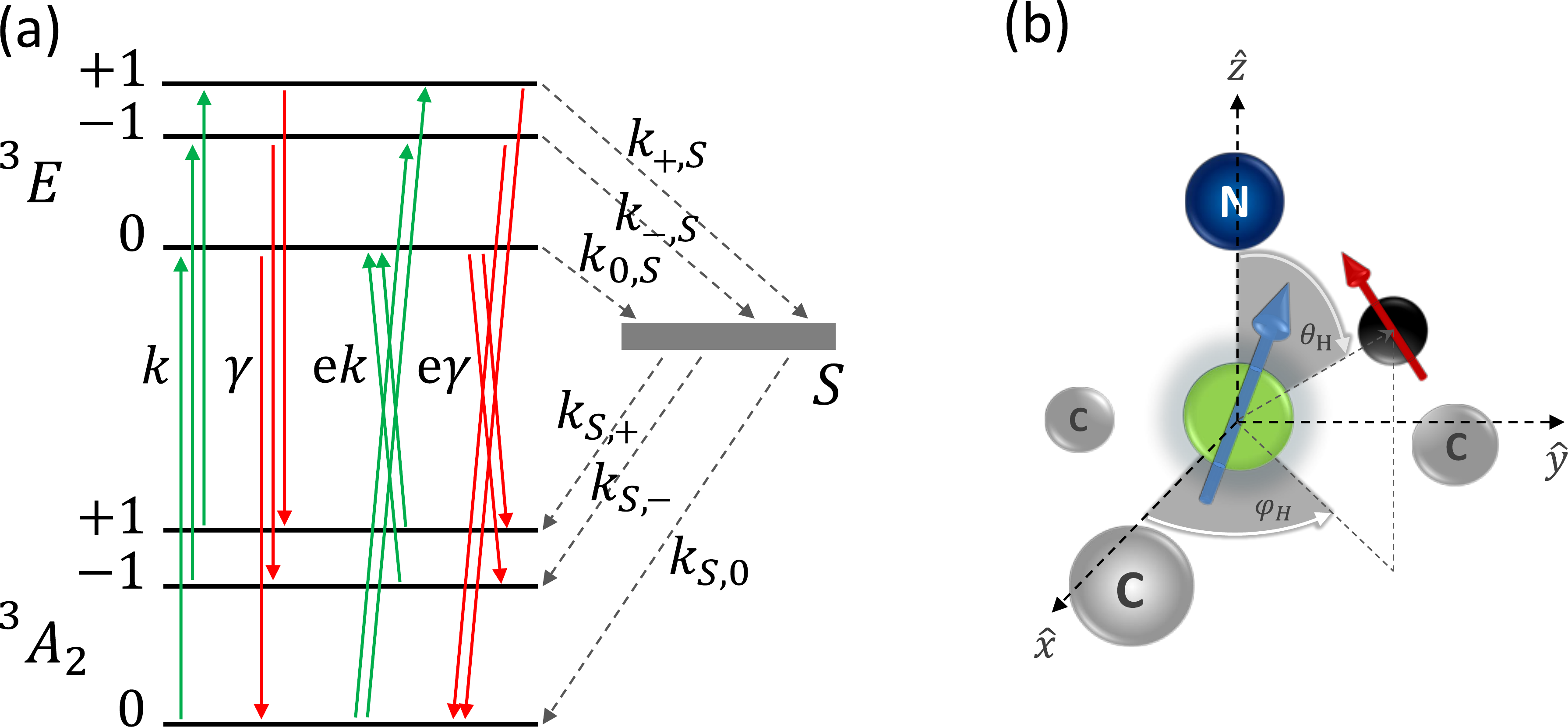}
    \caption{(a) Schematic representation of the seven-level model used in this study showing transition rates associated to the optical excitation ($k$, green arrows), spontaneous decay ($\gamma$, red arrows), crossing transitions due to spin mixing ($e k$ and $e \gamma$), and inter-system crossing transitions (dashed arrows). (b) Atomic configuration of the nitrogen-vacancy center representing the electronic spin (blue arrow) and a nuclear spin (red arrow), making a polar angle $\theta_H$ relative to the NV axis, $z$ axis, and an azimuthal angle $\varphi_H$ with respect to the $x$ axis which is in the plane that contains the NV axis and one of the three carbon atoms adjacent to the vacancy.}
     \label{fig_NV}
\end{figure}

\section{Electronic spin polarization}\label{secElectron}
\begin{table*}[t]
	\caption{\label{tabRates}Transition rates, based on Refs. \cite{Manson, Robledo, Tetienne, Gupta}, for the spontaneous decay $\gamma$ and inter-system crossing from different spin projections of the excited state to the single, $k_{m_s,S}$, and from the singlet to the ground state spin projections, $k_{S,m_s}$. The parameter $e$ is used to allow for optical transitions that do not preserve spin. See also Fig.~\ref{fig_NV}.}\label{Table_rates}
	\begin{ruledtabular}
		\begin{tabular}{ccccc}
			& Model 1 \cite{Manson}& Model 2 \cite{Robledo}& Model 3 \cite{Tetienne} & Model 4 \cite{Gupta} \\
			\hline
			$\gamma$ (MHz) & 77 & 62.7 & 63.2 & 67.4\\
			e & 1.5/77 & 0.01 & 0 & 0\\
			$k_{0,S}$ (MHz) & 0 & 12.97 & 10.8 & 9.9\\
			$k_{+,S}$ (MHz) & 30 & 80 & 60.7 & 91.6\\
			$k_{-,S}$ (MHz) & 30 & 80 & 60.7 & 91.6\\
			$k_{S,0}$ (MHz) & 3.3 & 3.45 & 0.8 & 4.83\\
			$k_{S,+}$ (MHz) & 0 & 2.16/2 & 0.4 & 2.11/2 \\
			$k_{S,-}$ (MHz) & 0 & 2.16/2 & 0.4 & 2.11/2 \\
		\end{tabular}
	\end{ruledtabular}
\end{table*}
The current understanding for the electronic spin polarization of NV centers in diamond is that upon optical illumination the electronic spin can be predominantly pumped into the $m_s=0$ state. Initially, it was proposed that only the electron in its $m_s=\pm1$ spin projections undergoes an inter-system crossing by transiting from the excited state $^3E (ae)$ with spin projections $m_s=\pm1$ to the singlet states (S) and from the singlet to the ground state ${^3A_2}(e^2)$ with spin projection $m_s=0$ \cite{Manson}. We denote these transitions by the rates $k_{\pm,S}$, and $k_{S,0}$, respectively (see Fig.~\ref{fig_NV}). As the optical transitions are taken to be mostly spin conserving, electronic spin will be mainly polarized in the $m_s=0$ state after few optical cycles. 

However, recent experiments have shown that additional rates must be included in the inter-system crossing \cite{Tetienne, Robledo, Gupta, Hacquebard}. Those experiments showed that electrons with spin projection $m_s=0$ on the excited state can also undergo the inter-system crossing with rate $k_{0,S}$. 
In addition, and more crucially, electrons can also relax from the singlet to the $m_s=\pm1$ spin projections with  rates $k_{S,\pm}$. This has important consequences on the electronic spin polarization, especially at large optical powers. 

In addition to the inter-system crossing rates, non-spin preserving optical transitions exist due to spin mixing caused by an intrinsic spin-spin interaction and magnetic field components perpendicular to the NV axis. We model the effect of the intrinsic mixing with parameter $e$ (see Fig.~\ref{fig_NV}) while the spin mixing caused by magnetic fields can be modeled separately. This spin mixing increases the population of the singlet, especially at large  optical powers, as the singlet population relaxes to the ground state at a rate which is about 30 times smaller than the spontaneous decay rate, denoted by $\gamma$. 

\begin{figure}[t]
\vspace*{-0.15cm}
\centering
    \includegraphics[width=.475\textwidth]{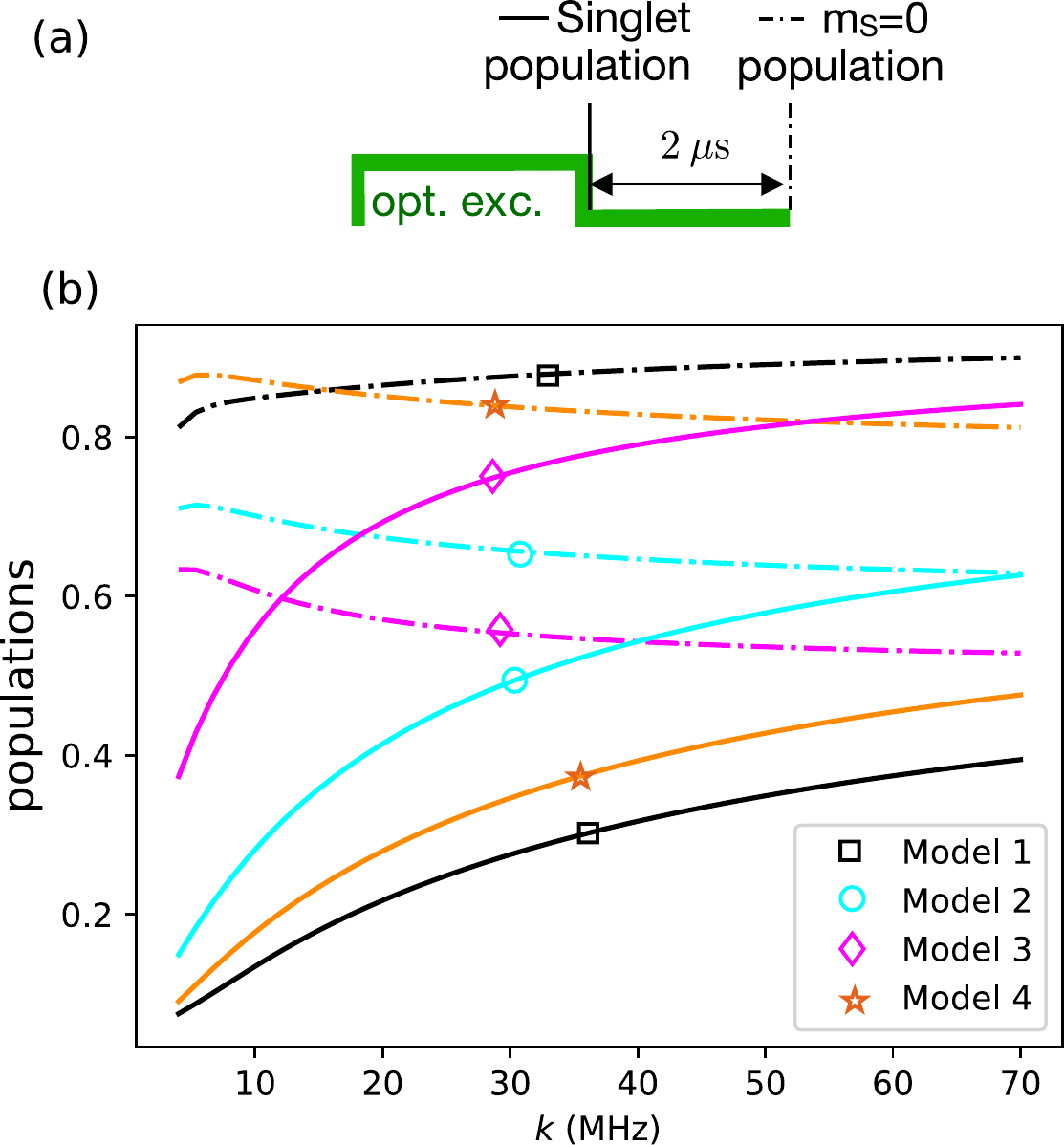}
   \caption{(a) Sequence for calculating the population of the singlet (after 2~$\mu$s of optical excitation) and the $m_s=0$ ground state of the electronic spin (after 2 $\mu$s waiting time after excitation so that singlet population relaxes to the ground state). (b) The singlet population (solid lines) and the $m_s=0$ ground state population (dash-dotted lines) as a function of the optical excitation rate, $k$, for the models given in Table \ref{Table_rates}. }
   \label{fig_ms0singlet}
\end{figure}

The models under our consideration are summarized in Table \ref{tabRates}. Model 1, adapted from Ref.~\cite{Manson}, has no transition rate from the singlet to $m_s=\pm1$ ground states ($k_{S,\pm}=0$). Therefore, as the optical excitation rate, labeled by $k$, increases, so does the electronic spin polarization (see Fig.~\ref{fig_ms0singlet}). However, model 1 does not result in complete electronic polarization because of non-spin conserving transition rates with $e=0.019$. Models 2, 3 and 4, adapted from Refs.~\cite{Robledo}, \cite{Tetienne} and \cite{Gupta}, respectively, have nonzero $k_{S,\pm}$ with model 4 having the largest ratio  $k_{S,0}/k_{S,\pm}$.  For these three models,  the electronic spin polarization decreases as the optical excitation rate $k$ increases. Note that, at low $k$ model 4 gives the highest electronic polarization, because of large ratio $k_{S,0}/k_{S,\pm} \approx 4.6$ and $e=0$. However, at large $k$, model 1 gives the highest electronic polarization. We note that, these models are taken at room temperature. The inter-system crossing rates may depend on temperature (see Ref.~\cite{Kalb}).

Figure \ref{fig_ms0singlet} also shows the singlet population for all models after 2 $\mu s$ of optical excitation. Transitions from excited states to the singlet populates the singlet state, which increases with the optical excitation rate. Therefore, the electronic spin polarization onto $m_s=0$ predominantly depends on the rate between $k_{S,0}$ and $k_{S,\pm}$, i.e. on $k_{S,0}/(k_{S,0}+k_{S,\pm})$. 

It is worth mentioning a few details about the transition rates we are using in this work. First, Refs.~\cite{Robledo}, and \cite{Gupta}, from which we have adapted models 2 and 4, respectively, use a five-level model for the electron spin of the NV center, i.e., $m_s=+1$ and $m_s=1$ states are assumed to be degenerate and is taken as one state, $m_s=\pm 1$. Here we use a seven-level model for the NV center. Therefore, for models 2 and 4 we have divided by 2 the transition rate given from the singlet to the $m_s=\pm 1$ states. Second, Refs.~\cite{Robledo}, \cite{ Tetienne}, and \cite{Gupta} have measured different rates for different NV centers. We have used the set of rates that give the highest electronic and therefore nuclear polarization. 

In the following section, we discuss how these transition rate models affect the nuclear spin polarization for several polarization methods. 

\section{Nuclear spin polarization methods}\label{secNuclear}
Nuclear spins can be polarized using the hyperfine interaction between the electronic and nuclear spins in several ways. The Hamiltonian for an NV electronic spin and a nuclear spin $1/2$ is given by~($\hbar=1$)
\begin{equation}\label{eq_H}
H_{i} = D_{i}S_z^2 + \gamma_{el} \mathbf{B}\cdot \mathbf{S} + \gamma_n \mathbf{B}\cdot \mathbf{I} + H_{i,\textrm{hf}}
\end{equation}
where $i=g,e$ denotes the ground and excited electronic states, $D_g/(2\pi)=2.87$~GHz ($D_e/(2\pi)=1.42$~GHz) is the ground (exited) zero field splitting between the $m_s=0$ and $m_s=\pm1$ spin states, and $\gamma_{el}/(2\pi)=2.8$ MHz/G is the electronic gyromagnetic ratio \cite{Neumann09}. The second and third terms are the electronic and nuclear Zeeman interactions, respectively. The gyromagnetic ratio of $^{13}$C, $^{14}$N and $^{15}$N nuclear spins, $\gamma_n/(2\pi)$, are 1.07 kHz/G, 0.3077 kHz/G, and -0.4316 kHz/G, respectively. The fourth term is the hyperfine interaction Hamiltonian given by
\begin{eqnarray}\label{eq_H_hpf}
H_{i,\textrm{hf}} &=& A^{(i)}_{zz}S_zI_z +\frac{A^{(i)}_{\perp}}{4}\left(S_+I_- + S_-I_+\right) \nonumber\\
&&+\frac{A'^{(i)}_{\perp}}{4}\left(S_+I_+e^{-2i\varphi_H} + S_-I_-e^{2i\varphi_H}\right) \nonumber\\
&&+\frac{A^{(i)}_{ ani}}{2}\left[\left(S_+I_z+S_zI_+\right)e^{-i\varphi_H}\right. \nonumber \\
&&+\left. \left(S_-I_z+S_zI_-\right)e^{i\varphi_H}\right].
\end{eqnarray}
in which $\varphi_H$ is the azimuthal angle (see Fig.~\ref{fig_NV}(b)). The parameters in the above equation depend on the relative position between the electronic and nuclear spins as follows
\begin{eqnarray}
&&A^{(i)}_{zz} = A_c - A_d(1-3\cos^2\theta_H),\\
&&A^{(i)}_{\perp}=2A_c + A_d(1-3\cos^2\theta_H),\\
&&A^{(i)}_{ ani}=3A_d \cos\theta_H\sin\theta_H, \label{eq_Aani}\\
&&A'^{(i)}_{\perp}=3A_d\sin^2\theta_H,
\end{eqnarray}
where $A_c$ is the contact term contribution which decays exponentially with distance between the electron and nuclear spins, $A_d$ is the dipole-dipole hyperfine coupling, decaying as $1/r^3$ for far nuclear spins \cite{Gali,Nizovtsev_NJP18}, and $\theta_H$ is the polar angle of the nuclear spin relative to the NV axis (see Fig.~\ref{fig_NV}(b)). Note that, in general, the hyperfine matrices are different for the ground and excited electronic states. It is straightforward to obtain the hyperfine Hamiltonian, given in Eq.~\eqref{eq_H_hpf}, from a description in Cartesian coordinates \cite{supplement}.

It is through the hyperfine interaction and the inter-system crossing mechanism that the nuclear spin can be polarized from thermal equilibrium. The first term in $H_{i,\textrm{hf}}$ can be considered as an energy shift of the electronic spin depending on the nuclear spin state. The second term causes spin flip-flops between the electronic and nuclear spins when this process nearly preserves energy. The third term represents non-energy preserving spin flips. The last term represents rotation of either the electron or the nuclear spin without rotating the other spin. Some of these terms can lead to nuclear spin polarization depending on the external optical excitation, magnetic field and state of the electronic spin. 

Next, we describe the three methods for polarizing a nuclear spin. In our simulations we consider a nucleus with spin $1/2$, i.e., $^{13}$C or $^{15}$N nuclear spin. We label the basis states as $|m_s,m_I\rangle$, where the first component indicates the electron spin projection, $m_s=0,\pm 1$, and the second component determines the nuclear spin projection $\uparrow(\downarrow)$ corresponding to $m_I=\pm 1/2$, respectively. The discussion is also valid for $^{14}$N nuclear spin 1. We note that, in the case of $^{14}$N, the Hamiltonian has the extra term $Q I^2_z$, for the quadrupole interaction of the nuclear spin in which $Q=-4.96$ MHz \cite{Busaite}.

\subsection{ESLAC}\label{secESLAC}
\begin{figure*}[th]
\centering
    \includegraphics[width=1\textwidth]{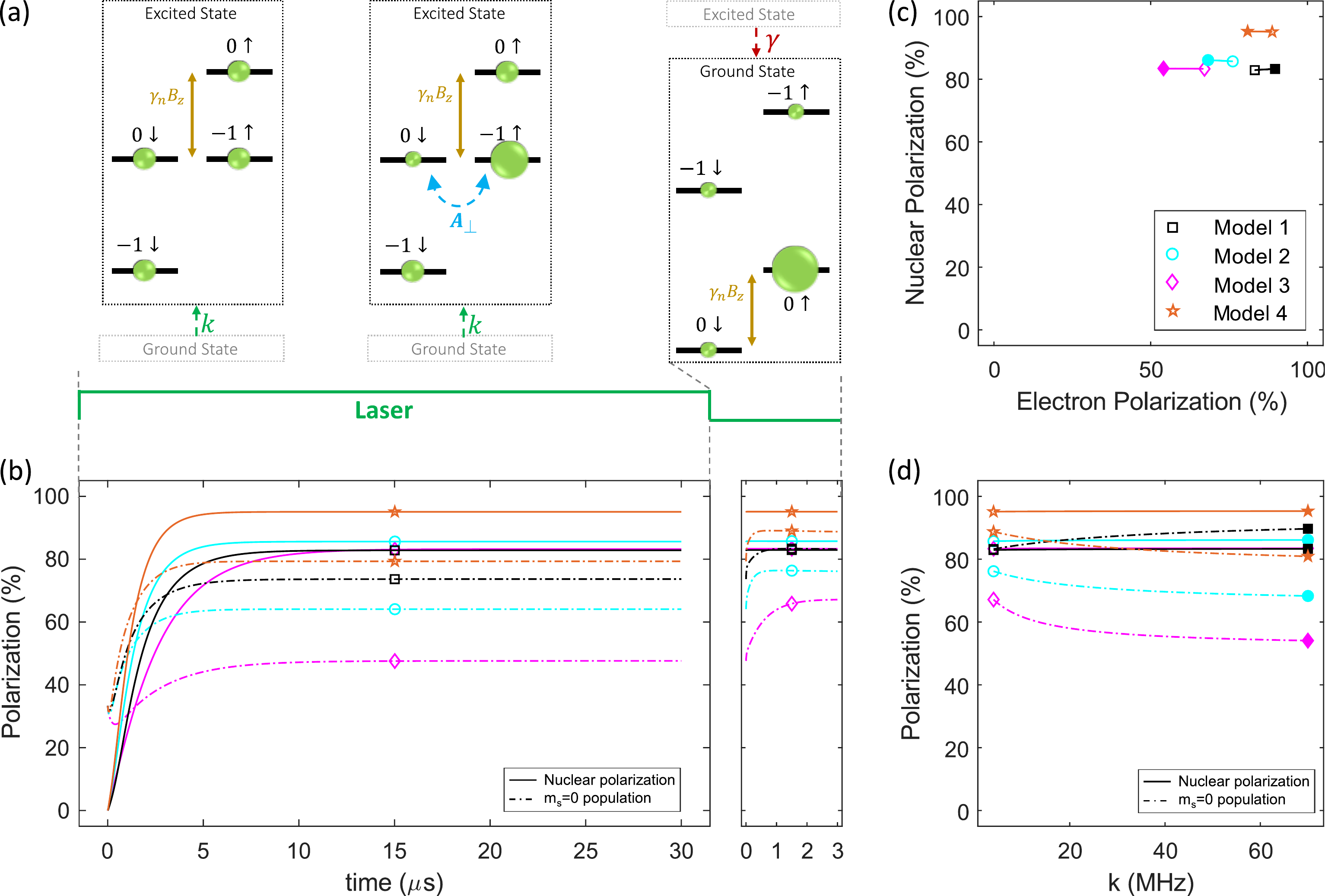}
    \caption{ \label{figESLAC} (a) A diagram showing the population of the eigenstates of $S_z$ and $I_z$ which are mainly involved in nuclear polarization at ESLAC during the laser and waiting time. The dashed arrows, labeled by $k$ and $\gamma$, represent the optical excitation  and the spontaneous decay, respectively. (b)  Electronic and $^{15}$N nuclear spin polarization dynamics for $k=4$ MHz for the four transition rate models given in Table \ref{Table_rates} as in the legends of (c). (c) Nuclear versus electronic spin polarizations parametrized by the excitation rate $k$. The open markers correspond to $k = 4$ MHz, while the filled markers correspond to $k = 70$ MHz. (d) Electronic and nuclear polarization as a function of $k$.} 
\end{figure*}

In this approach, an external magnetic field, $\approx 510$ G, is applied along the NV axis ($z$ axis) so that the energy levels $m_s=-1$ and $m_s=0$ in the excited state become very close (see Figure \ref{figESLAC}(a)). Under this configuration, the second term of the hyperfine Hamiltonian, Eq.~\eqref{eq_H_hpf}, mixes states $|0,\downarrow\rangle$ and $|-1,\uparrow\rangle$ in the excited state. Note that $m_s=1$ states ($|1,\downarrow\rangle$ and $|1,\uparrow\rangle$) are not mixed with $m_s=0$ and $m_s=-1$ states because they are far away in energy.

This scheme polarizes the nuclear spin to $|\uparrow\rangle$, as we now explain \cite{Vincent}. The optical excitation transfers the electron from $|0,\downarrow\rangle$ ground state to $|0,\downarrow\rangle$ excited state. This is shown by green arrows labeled by $k$ in Fig.~\ref{figESLAC}(a). In the excited state, the $A^{(e)}_{\perp}$ component in~$H_{e,hf}$ causes precession between $|0,\downarrow\rangle$ and $|-1,\uparrow\rangle$ (shown by the blue arrow in Fig.~\ref{figESLAC}(a)). During this precession the electronic spin may go to $|0,\uparrow\rangle$ in the ground state by passing through the meta-stable singlet state (shown by the red arrow labeled by $\gamma$).  This polarization method relies primarily on the rates from the excited state to the singlet and secondarily on the rates from the singlet to the ground state.

Figure \ref{figESLAC} shows the nuclear spin polarization for $^{15}$N. The hyperfine matrix for this nuclear spin is diagonal for both ground and excited states ($A_{ ani}=0$) and are given in Ref.~\cite{Ivady} (see Table \ref{tabAmat}). In Fig.~\ref{figESLAC}(a) we show the sequence for polarizing the nuclear spin. The dynamics of the nuclear and electronic polarizations are shown in Fig.~\ref{figESLAC}(b). We indicate the electron polarization by the population in $m_s=0$ state, while for the nuclear polarization we use  
\begin{equation}
P_{ n}=|\rho_n(\uparrow,\uparrow)-\rho_n(\downarrow,\downarrow)|,
\end{equation}
where $\rho_n$ is the reduced density matrix after tracing over the electronic spin. 

We have calculated the density matrix evolution using the master equation \cite{Havel, supplement} assuming no initial polarization for both the electronic and nuclear spins. After few microseconds of optical excitation, the nuclear spin is polarized at a rate proportional to $A^{(e)}_{\perp}$. For more details see Refs.~\cite{Vincent, GaliPRB09}. In our simulations we have taken the transverse and longitudinal relaxation times of the NV electron spin as $T^{\star}_{2,el}=3$ $\mu$s (6 ns) for the ground (excited) state and $T_{1,el}=1$ ms, respectively, and $T^{\star}_{2,n}=1$ ms and $T_{1,n}=100$ ms for the nuclear spin. We have also neglected ionization effects of the NV center due to optical excitation \cite{Poggiali}.

As expected, once the optical excitation is turned off the optical excited and singlet populations relax to the ground state, increasing the electronic polarization. As the optical excitation rate increases, the nuclear polarization remains almost unchanged while the electronic polarization increases slightly for model 1 and decreases for the other models (see Fig.~\ref{figESLAC}(c) and (d)). Figure~\ref{figESLAC}(c) shows the nuclear polarization achieved for a specific electronic polarization where polarizations are parametrized by the optical excitation rate, $k$. The empty markers correspond to $k=4$ MHz while the filled markers correspond to $k=70$ MHz.

Non-zero rates $k_{0,S}$, $k_{S,\pm}$, and non-spin conserving transitions, parametrized by $e$, will result in increasing the population of $m_s=\pm1$ on the ground state, which will contribute to the polarization of the opposite nuclear spin projection. Therefore, the model for which $k_{\pm,S}/k_{0,S}$ and $k_{S,0}/k_{S,\pm}$ are larger and $e$ is smaller, result in a higher nuclear polarization. Figure \ref{figESLAC} shows that model 4 gives the highest $^{15}$N nuclear polarization, $\approx 95\%$, at ESLAC, very close to the experimentally observed value 96$\%$ \cite{Vincent}. Note that, although model 1 has  $k_{0,S}=k_{S,\pm}=0$,   non-spin conserving transitions in this model contribute to depolarization of the nuclear spin, resulting in a lower nuclear polarization.
In the supplementary materials \cite{supplement} we compare our simulations for the $^{15}$N nuclear polarization for a range of magnetic fields with the experimental data of Ref.~\cite{Vincent}. 

\subsection{Polarization by nuclear spin precession on $m_s=0$}\label{secMS0}
 \begin{figure*}[t!]
\centering
    \includegraphics[width=1\textwidth]{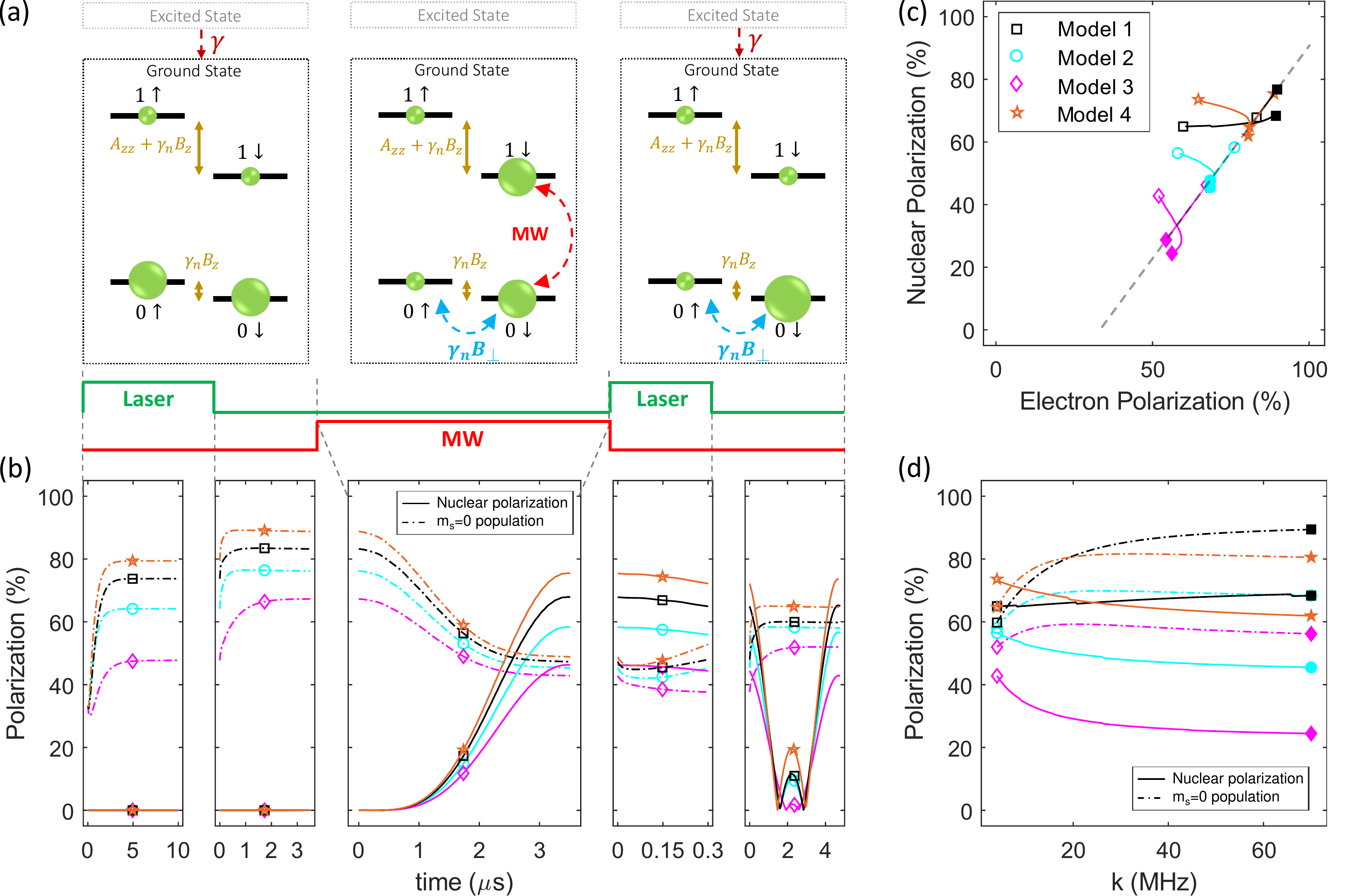}
    \caption{ \label{figMS0}  (a) Eigenstates of $S_z$ and $I_z$ involved in the polarization based on nuclear precession on $m_s=0$. The green circles show the populations during the laser and microwave (MW) sequences. The red dashed arrow $\gamma$, represents the spontaneous decay to the ground state. (b) Electronic  and nuclear spin polarization dynamics for $k=4$ MHz for the four transition rate models. (c) Nuclear versus electronic spin polarization parametrized as a function of the optical excitation rate $k$. Open (filled) markers correspond to $k= 4$ MHz ($70$ MHz). The data on the dashed-gray line correspond to the electron spin polarization after the first waiting time and nuclear spin polarization after the MW. The data that does not sit on the dashed-gray line correspond to the polarizations at the end of the sequence. (d) Electronic and nuclear polarization at the end of the sequence as a function of $k$. This figure is plotted for a $^{13}$C nuclear spin in family C (see Section \ref{secMS0}).}
    \end{figure*}
In this method, the electron spin is first optically pumped. After waiting for the population to decay to the ground state, indicated by the red arrow labeled by $\gamma$ in Fig.~\ref{figMS0}(a), a selective microwave pulse is applied in order to transfer the population from $|0, \downarrow\rangle$ to $|1, \downarrow\rangle$. Meanwhile, the nuclear spin precesses between $|0,\uparrow\rangle$ and $|0,\downarrow\rangle$, due to a perpendicular magnetic field (indicated by the blue arrow in Fig.~\ref{figMS0}(a)). 
Finally a laser pulse, followed by a waiting time, is used to leave the electron spin mostly in its $m_s=0$ ground state (final red arrow labeled by $\gamma$ in Fig.~\ref{figMS0}(a)). 

This approach is a variation of the method proposed in Ref.~\cite{Dutt}. In that work, a selective microwave $\pi$ pulse is used, while here microwave excitation and nuclear precession take place simultaneously \cite{supplement}. Optimizing the  microwave time results in a higher nuclear polarization. This time is given by $ t_{mw} \approx \pi/(\sqrt{2}|\Delta|)$ where $\Delta$ is the coupling between $|0,\uparrow\rangle$ and $|0,\downarrow\rangle$ \cite{supplement}. This approach can be also understood by means of coherent population trapping (see Ref.~\cite{Raul} for details).

\begin{table}[b]
	\caption{\label{tabAmat} The ground state (GS) and excited state (ES) hyperfine components (in MHz) for $^{15}$N and families C and H of carbon nuclear spins.}
	\begin{ruledtabular}
		\begin{tabular}{ccccc}
			family, state & $A_{zz}$ & $A_{ani}$ & $A_\perp$ &  $A_{\perp}'$  \\
			\hline
			$^{15}$N, GS & 3.4 & 0 & 7.8 & 0 \\
			$^{15}$N, ES & -58.1 & 0 & -77 & 0 \\
			C, GS & -8.822 & -0.789 & -20.378 & 0.621 \\
			C, ES & -3.78  & 0.749  & -14.12 & 0.680 \\
			H, GS & 1.933 & -0.250 & 2.067 & 0.670\\
			H, ES &  3.413 & -0.349  & 4.086 & 0.866\\
		\end{tabular}
	\end{ruledtabular}
\end{table}
Figure \ref{figMS0}(b) shows the dynamics of the NV electronic spin and a  $^{13}$C nuclear spin in family C for a perpendicular magnetic field component of $B_x=10$~G and a small parallel component of $B_z=0.5$~G. The classification of nuclear spins to families is proposed in Refs.~\cite{Smeltzer, Dreau} (see Ref.~\cite{Smeltzer} for a diagram of $^{13}$C families in the diamond lattice). The hyperfine components of this nuclear spin is given in Table \ref{tabAmat}. The highest nuclear spin polarization is achieved after the microwave pulse and is proportional to the electron spin polarization achieved due to the first optical excitation. Figure~\ref{figMS0}(c) shows the electron and nuclear polarizations parametrized by the optical excitation rate, $k$. Similar to the ESLAC case, the empty markers correspond to $k=4$ MHz and the filled markers correspond to $k=70$ MHz. This method crucially depends on the electronic spin polarization. This dependence can be clearly observed by comparing the nuclear polarization at the end of the microwave with the electronic polarization at the end of the first waiting time, shown by the points on the dashed gray line. Here, also the highest nuclear polarization is achieved for model 4. Model 1 shows higher nuclear polarization at larger optical excitations. However, models 2 to 4 show the opposite behavior.  

The points that do not sit on the dashed gray line are polarizations at the end of the sequence, and show how the nuclear spin depolarizes under optical excitation \cite{Jiang}. We have chosen the time of the second laser pulse, 300 ns, in such a way that we achieve polarization for both the electron and the nuclear spin at the end of the sequence. The second waiting time is chosen sufficiently large so that the nuclear spin can be recovered due to its precession, i.e., $\approx 1/(2|\Delta|)$. It can be noted that during the second waiting time, the direction of the nuclear polarization changes due to the precession of the nuclear spin about the external magnetic field $B_x$. 
 
Figure~\ref{figMS0}(d) shows the nuclear and electronic polarizations at the end of the sequence as a function of the excitation rate $k$. Due to the short time of the second laser pulse, as $k$ increases, the electronic polarization increases for all models at small $k$. On the other hand, the nuclear polarization increases for model 1, while it decreases for the other models. The second optical excitation results in depolarization of the nuclear spin. However, after the first waiting time, increasing $k$ results in a higher (lower) electronic polarization for model 1 (model 2-4) and therefore a higher (lower) nuclear polarization after the microwave. 
This observation together with an experimental realization of the nuclear polarization as a function of the optical power can be used to test the validity of the transition rate models. Note that, for this method, the second highest nuclear polarization is achieved with model 1, as opposed to the ESLAC method where the second highest is achieved by model 2. This is because the non-spin conserving transition rates in model 1 is higher than model 2 which result in depolarization of the nuclear spin in ESLAC.

In our simulations, we have used the secular approximation, i.e., we have only kept terms in the Hamiltonian (Eq.~\eqref{eq_H}) proportional to $S_z$ and have ignored the terms that contain $S_x$ or $S_y$. We have taken the effect of the non-secular terms perturbatively by adding the following Hamiltonian \cite{Jero_PRB08, Childress_Science, supplement}
\begin{eqnarray}\label{eq_Hsoc}
H_{\rm soc}&&=\frac{(3S^2_z-2)D+S_z\gamma_{el} B_z}{2(D^2-\gamma^2_{el} B^2_z)}\hat M\nonumber \\
&&\quad+\frac{(2-S^2_z)\gamma_{el} B_z-S_z D}{2(D^2-\gamma^2_{el} B^2_z)}\hat N,
\end{eqnarray}
where
\begin{eqnarray}
\hat M&&= 2\gamma_{el}\left[\left(A_{xx}B_x+A_{yx}B_y\right) I_x \right. \nonumber \\ 
&&\quad+ \left(A_{xy}B_x+A_{yy}B_y\right)I_y \nonumber\\
&&\quad+\left. \left(A_{xz}B_x+A_{yz}B_y\right) I_z\right]\nonumber \\
&&\quad+ \gamma^2_{el} B^2_{\perp}\mathbf{1}+(\vec A_{+}\cdot \vec A_{-})\mathbf{1},
\end{eqnarray}
and 
\begin{eqnarray}
\hat N= i (\vec A_{+}\times \vec A_{-})\cdot \vec I.
\end{eqnarray}
Here, $\vec I$ is a vector which its components are the nuclear spin matrices $\vec I=(I_x, I_y, I_z)$, and $\mathbf{1}$ is the $2\times 2$ identity matrix.

Note that $|\uparrow\rangle$ and $|\downarrow\rangle$ are the eigenstates when the electronic spin is in $m_s=1$ of the ground state. For $m_s=0$ electron spin, the hyperfine Hamiltonian is zero and the off-axis magnetic field (shown by the blue arrow in Fig.~\ref{figMS0}(a)) together with the correction given in Eq.~\eqref{eq_Hsoc} determine the quantization axis of the nuclear spin. The precession rate is given by $\gamma_n B \left(1-|\hat{n}_{m_s=0}\cdot \hat{n}_{m_s=1}|\right)$ where $\hat{n}_{m_s}$ is the unit vector that indicates the quantization axis of the nuclear spin while the electron is in the $m_s$ state. $\hat{n}_{m_s=0}$ is mostly determined by the external magnetic field while $\hat{n}_{m_s=1}$ is mostly determined by $\mathbf{A}_z+\gamma_n \mathbf{B}$. There is an optimal perpendicular magnetic field that results in a higher polarization at the end of the sequence. If the magnetic field is too low, it results in a lower polarization after the microwave. While if the magnetic field is too high, both quantization axes are similar and the precession does not take place. Nuclear spins with diagonal hyperfine matrices and/or high contact term, i.e. small $A_{ani}$ component, can be polarized with this method as we will discuss in Section~\ref{secTheta}.
 
 \begin{figure*}[t!]
\centering
    \includegraphics[width=0.95\textwidth]{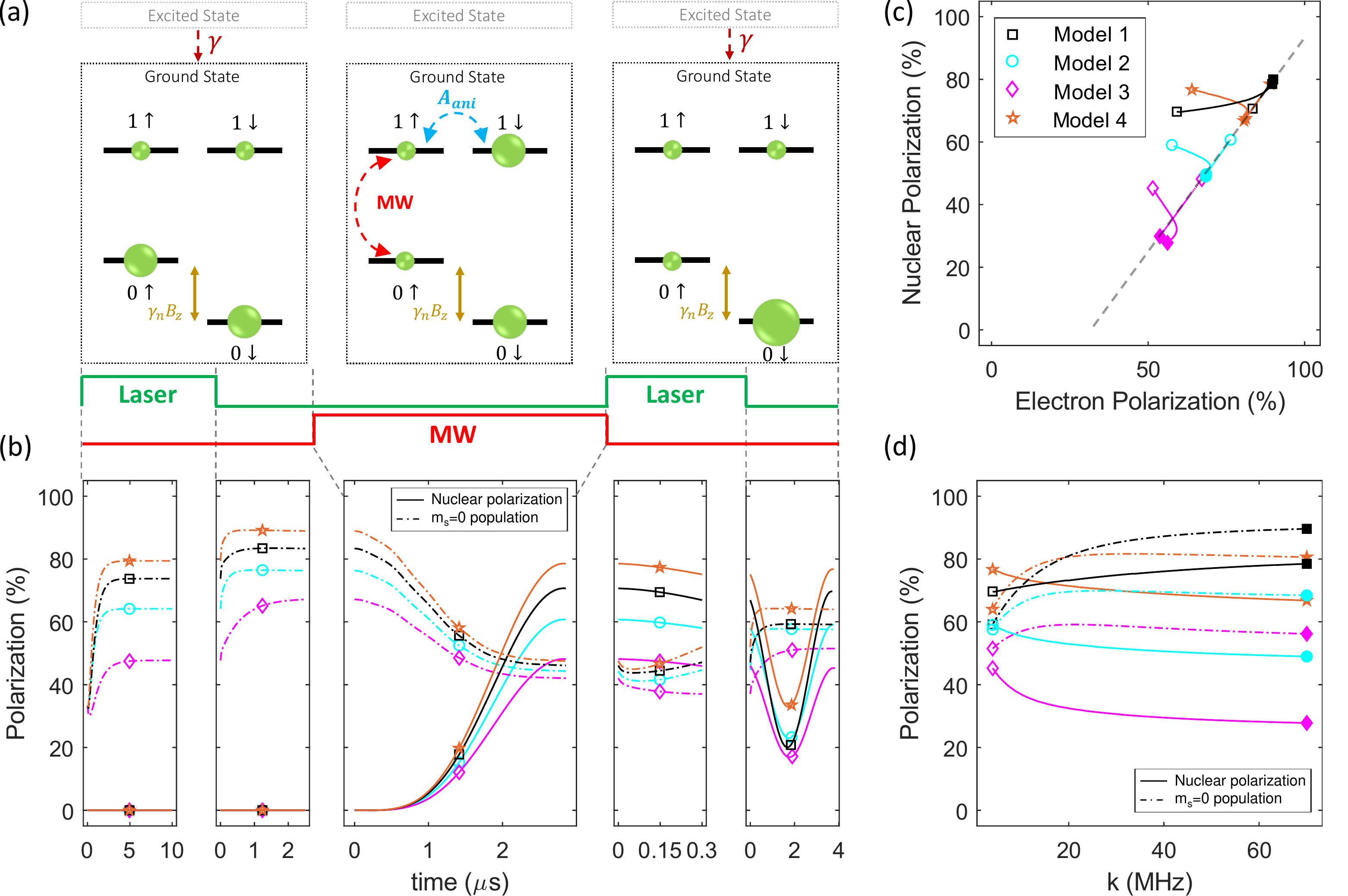}
    \caption{ \label{figMS1} (a) The eigenstates of $S_z$ and $I_z$ which are mainly involved in the polarization based on nuclear spin precession on $m_s = 1$. The green circles show the population of the states during the laser and microwave pulses. The red dashed arrow, $\gamma$, represents the spontaneous decay. (b) Electronic and nuclear spin polarization dynamics for $k=4$ MHz for the four transition rate models. (c) Nuclear versus electronic spin polarization parametrized by the optical excitation rate $k$. The open (filled) markers correspond to the optical excitation rate $k = 4$ MHz (70 MHz).  The data on the gray dashed line correspond to the electron spin polarization after the first waiting time and nuclear spin polarization after the MW. The data that does not sit on the dashed-gray line corresponds to the polarizations at the end of the sequence. (d) Electronic and nuclear polarization at the end of the sequence as a function of $k$. In this figure we have considered a $^{13}$C nuclear spin in family H.}
\end{figure*}

We finish this section by noting that the nuclear polarization can be improved by repeatedly applying the sequence in Fig.~\ref{figMS0}(a) \cite{Dutt}.
 
\subsection{Polarization by nuclear spin precession on $m_s=1$}\label{secMS1}
Nuclear spins with $A_{ani}\neq 0$ can be polarized by using their precession while the electronic spin is in the $m_s=1$ state \cite{Jamonneau_thesis}. In this case, it makes sense to consider a nuclear spin basis given by the eigenstates while the electronic spin is in the $m_s=0$ state. The precession relies on the anisotropic component of the hyperfine interaction, which depends on the relative orientation of the nuclear spin with respect to the NV axis (see Eq.~\eqref{eq_Aani}). This precession effectively takes place if we apply a magnetic field along the NV axis that brings the ground states $|1,\uparrow\rangle$ and $|1,\downarrow\rangle$ close, i.e., $B_z=-A_{zz}/\gamma_n$.

For nearby nuclear spins, which have $A_{zz}$ of the order of few MHz, this method requires very large magnetic fields, of the order of few thousand Gauss. Therefore, this method may be experimentally more accessible for far $^{13}$C nuclear spins which have $A_{zz}$ of the order of few hundred kHz, e.g. for family J and families further away \cite{Nizovtsev_NJP14}, for which magnetic fields below 1000 G are required (see Fig.~\ref{fig_pn}(f)).

Figure \ref{figMS1}(a) illustrates a typical configuration of this protocol. After the optical excitation and spontaneous decay (indicated by the red arrow labeled by $\gamma$ in Fig.~\ref{figMS1}(a)), the electron is polarized to the $m_s=0$ ground state. Following that, a selective microwave excitation transfers the population from $|0,\uparrow\rangle$ to $|1,\uparrow\rangle$. Due to the anisotropic component of the hyperfine interaction, the nuclear spin precesses between the states $|1,\uparrow\rangle$ and $|1,\downarrow\rangle$. This process effectively transfers the electronic polarization to the nuclear spin polarization.

Figure \ref{figMS1}(b) shows the dynamics for a $^{13}$C nuclear spin in family H with its hyperfine components given in Table \ref{tabAmat}. It is clear that if the electronic polarization is not perfect, neither is the nuclear polarization. This effect can be seen in Fig.~\ref{figMS1}(c) where both polarizations are plotted parametrized by the excitation rate. Similar to the method based on precession in the $m_s=0$ state, data points on the dashed gray line correspond to the electronic polarization after the first waiting time and the nuclear polarization after the microwave. The data points that sit outside the dashed line correspond to the polarizations at the end of the sequence. Here, we have chosen the time of the second laser pulse and waiting time similar to the method based on precession on $m_s=0$ state. The discussion for figures \ref{figMS0}(c) and (d) is also relevant here. We note that, in our simulations we have used secular approximation, keeping only terms that contain $S_z$ in the Hamiltonian.


In the following Section we discuss the efficiency of the methods for different angular position of the nuclear spins.

\section{Anisotropy dependence of nuclear spin polarization}\label{secTheta}
Each of the discussed methods relies on different components of the hyperfine interaction, which in turn depends on the angular position of the nuclear spin relative to the NV axis. In this section we discuss the nuclear polarization performance as a function of the polar distribution of the nuclear spins relative to the NV axis. Moreover, we estimate the nuclear spin polarization for a range of families. 

\begin{figure*}[t]
\hspace*{-0.55cm}
\centering
 \includegraphics[width=1\textwidth]{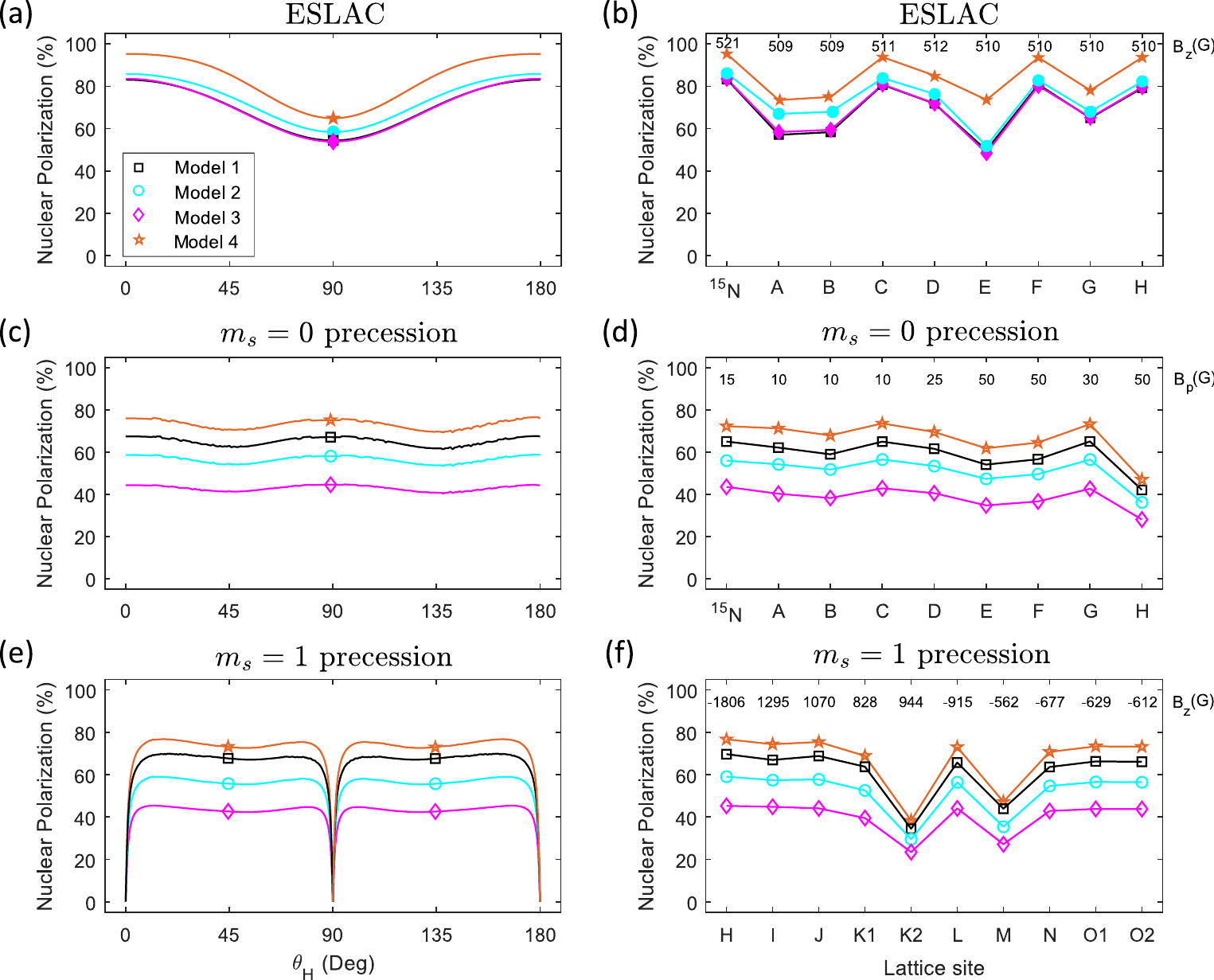}
    \caption{ \label{figTHETA} Nuclear polarization versus $\theta_H$ for the method based on ESLAC (a), precession in $m_s=0$ (c), and precession in $m_s=1$ (e). In these plots we have respectively used the contact and dipole terms corresponding to families C, C, and H, setting $\varphi_H=0$. In (b), (d) and (f) we show the nuclear polarization achieved with different methods for a range of families. In these figures, the upper horizontal axes show the magnetic field for each family.}
    \label{fig_pn}
\end{figure*}

The method based on ESLAC takes advantage of the perpendicular component of the hyperfine interaction, $A_\perp$, to cause flip-flops between the electron and nuclear spins. If $A_c>A_d$, $A_\perp$ is nonzero for all angles $\theta_H$. For $A_c\leq A_d$, there are two angles for which $A_\perp=0$. For zero contact term ($A_c=0$), we have $A_\perp=0$ if $\theta_H= \pm \arccos(1/\sqrt{3})\approx \pm 54\deg$. For far nuclear spins $A_\perp$ decreases, and as a result, the nuclear polarization rate decreases, requiring longer times to achieve polarization. On the other hand, the final nuclear polarization decreases as the $A^\prime_{\perp}$ and $A_{ ani}$ components of the hyperfine interaction increases \cite{supplement}. Therefore, the achieved nuclear polarization depends on the polar angle $\theta_H$ (Fig.~\ref{fig_pn}(a)). Figure \ref{fig_pn}(b) shows the nuclear polarization for $^{15}$N and families A to H of $^{13}$C, whose hyperfine matrices are taken from Ref.~\cite{Ivady}. The nuclear polarization is smaller for families that have $A_{ani}$ and $A'_{\perp}$ components comparable to $A_{\perp}$. In the supplementary materials we compare our simulated nuclear polarization for those families with the experimental data of Refs.~\cite{Vincent, Dreau, Smeltzer}. 

The methods based on nuclear precession while the electronic spin is in $m_s=0$, and $m_s=1$ depend on an external magnetic field perpendicular to the nuclear quantization axis $\mathbf{\Omega}_1=\mathbf{A}_z + \gamma_n \mathbf{B}$, where $\mathbf{A}_z = (A_{ani} \cos\varphi_H, A_{ani}\sin\varphi_H,A_{zz})$. In other words, defining $\mathbf{\Omega}_0=\gamma_n \mathbf{B}$, the magnetic field should be chosen such that
\begin{equation}\label{eq_Omega}
\mathbf{\Omega}_1 \cdot \mathbf{\Omega}_0=0.
\end{equation}

As we have taken $\varphi_H=0$ \cite{supplement}, a magnetic field that satisfies Eq.~\eqref{eq_Omega} is $\mathbf{B}=B \hat x=-(A_{ani}/\gamma_n)\hat x$. This perpendicular magnetic field could be very large for nuclear spins with $A_{ani}$ of the order of few hundred kHz. As this magnetic field is present during the whole sequence, we choose it smaller than $50$~G in order to achieve a high electronic polarization in the method based on precession on $m_s=0$. In addition, the magnetic field should be taken such that it does not change the quantization axis of the nuclear spin, i.e., $B\ll |\mathbf{A}_z|/\gamma_n$ when the electronic spin is in $m_s=1$. On the other hand, the precession rate on $m_s$ is proportional to $\Omega_{m_s}$, therefore even for $A_{ani} = 0$, a magnetic field larger than zero is needed for the polarization method based on $m_s=0$ precession.

Fig.~\ref{fig_pn}(c) shows a small dependence on $\theta_H$ because the correction term for the Hamiltonian (Eq.~\eqref{eq_Hsoc}) depends on non-diagonal elements of the hyperfine matrix, which in turn depend on $\theta_H$. In fact, a relatively lower nuclear polarization is achieved for family H, for which this correction reduces the nuclear spin precession (see Fig.~\ref{fig_pn}(d)).

For the method based on precesssion on $m_s=1$, the magnetic field is taken along the $z$ axis ($\mathbf{\Omega}_0=\Omega_0 \hat z$), and following Eq.~\eqref{eq_Omega}, is given by $B=-A_{zz}/\gamma_n$. Having the magnetic field along the NV axis, results in a higher electronic spin polarization as non-spin-preserving transitions are minimized. For this method, the precession between the nuclear spin states occurs due to the anisotropic component of the hyperfine interaction $A_{ani}$, which is zero at $\theta_H=\{0,\pi/2,\pi\}$ (see Fig.~\ref{fig_NV}). At these angles, no precession takes place and no nuclear spin polarization is achieved (see Fig.~\ref{fig_pn}(e)).

Figure~\ref{fig_pn}(f) shows the nuclear polarization for families H to O2. We have taken the hyperfine matrix for families I to O2 from Ref.~\cite{Nizovtsev_NJP14}. In that work, the hyperfine matrix is only given for the ground state of the NV center. As an approximation, we have used the same hyperfine matrix for the excited state. A lower nuclear polarization is achieved for families K2 and M. Family K2 has a small anisotropic term resulting in a low nuclear polarization. For family M, $A_{ani}$ is large enough that can cause transitions between $|0,\downarrow(\uparrow)\rangle$ and $|1,\downarrow(\uparrow)\rangle$, therefore, reducing the nuclear polarization. This method cannot be used to polarize the $^{15}$N nuclear spin as its hyperfine matrix is diagonal and the anisotropic term is zero.

As a summary, for nuclear spins close to the NV center, nuclear polarization can be achieved using methods based on ESLAC  and precession of electron spin while being on $m_s=0$ state. For such nuclear spins the method based on precession in $m_s=1$ requires very large magnetic fields and therefore is more susceptible to magnetic field misalignments. For far nuclear spins the methods based on $m_s=0$ and $m_s=1$ precession could be used. The ESLAC method for far nuclear spins requires a very long laser time and it will be limited by non-spin-preserving transitions.

 \section{Conclusions}
We have shown that the nuclear spin polarization might crucially depend on the transition rates involved in the inter-system crossing depending on the polarization method. In particular, the rates from the singlet to the $m_s=\pm 1$ spin projections cause a strong power dependent polarization of the electronic spin, which in turns affect the nuclear polarization. The nuclear polarization method which is less affected by the optical power is that based on the ESLAC for which the nuclear polarization changes very slightly for all the transition rate models. The other two methods that rely on nuclear spin precession on the ground state (either $m_s=0$ or $m_s=1$) are greatly affected by a finite electronic spin polarization. This is because during the precession, at most, the electron spin polarization is transferred to the nuclear spin polarization. 

Due to the lack of experimental data, we were only able to compare our simulations with the experimental data of ESLAC method. Using the experimental data of Ref.~\cite{Vincent} for this method, we showed that model 4 fits better to the experimental data.

We have also compared the polarization performance of these three nuclear polarization methods depending on the angular position of the nuclear spin relative to the NV axis. This analysis could give directions to achieve larger nuclear spin polarization and/or design experiments to further investigate the inter-system crossing rates using the nuclear spins as a measuring tool. Enhancing the polarization of nuclear spins could result in the enhancement of the NMR and magnetic resonance imaging. Moreover, it could enhance the coherence time of the NV electron spin.

\acknowledgements
The authors thank A. Dr\'eau for helpful discussions. H.D.~acknowledges support from Conicyt doctorado grant No.~21100070. H.T.D.~acknowledges support from the Fondecyt-postdoctorado grant No.~3170922 and Universidad Mayor through a postdoctoral fellowship. J.R.M.~acknowledges support from Fondecyt Regular Grant No.~1180673, Air Force grant number FA9550-18-1-0513, ONR grant number N62909-18-1-2180, and Anid PIA ACT192023. H.T.D., V.J. and J.R.M.~acknowledge support from Conicyt-ECOS grant C16E04.





\end{document}



\widetext
\begin{center}
\textbf{\large Supplemental Materials for Effect of inter-system crossing rates and optical illumination on the polarization of nuclear spins nearby nitrogen-vacancy centers}\label{SM}\newline

{{\hspace{2cm}H. Duarte$^{1,2}$, H. T. Dinani$^{1,3}$, V. Jacques$^4$, J. R. Maze$^{1,5}$}}\newline

$^{1}$Facultad de F\'isica, Pontificia Universidad Cat\'olica de Chile, Santiago 7820436, Chile \newline
$^{2}$ Faculty of Physics, Universidad T\'ecnica Federico Santa Mar\'ia, Avda.~Vicu\~{n}a Mackenna 3939, Santiago, Chile \newline
$^{3}$Centro de Investigaci\'{o}n DAiTA Lab, Facultad de Estudios Interdisciplinarios, Universidad Mayor, Santiago, Chile \newline
$^4$L2C, Laboratoire Charles Coulomb, Université de Montpellier and CNRS, Montpellier 34095, France \newline
$^5$Research Centre for Nanotechnology and Advanced Materials, Pontificia Universidad Cat\'olica de Chile, \\ 
\hspace{5cm}Santiago, Chile\newline
\end{center}
\setcounter{equation}{0}
\setcounter{figure}{0}
\setcounter{table}{0}
\setcounter{section}{0}
\makeatletter
\renewcommand{\theequation}{S\arabic{equation}}
\renewcommand{\thefigure}{S\arabic{figure}}
\renewcommand{\bibnumfmt}[1]{[S#1]}
\renewcommand{\citenumfont}[1]{S#1}
\section{Hyperfine interaction Hamiltonian} 
In this section we show how the hyperfine Hamiltonian can be written as given in Eq.~(2) of the main text. The hyperfine part of the Hamiltonian in tensor form is given as $\textbf{S}\cdot\textbf{A}\cdot\textbf{I}$ which can be written as 
\begin{eqnarray}
\textbf{S}\cdot\textbf{A}\cdot\textbf{I}&&= S_xI_xA_{xx} + S_xI_yA_{xy} + S_xI_zA_{xz}\nonumber\\
&&\quad+ S_yI_xA_{yx} + S_yI_yA_{yy} + S_yI_zA_{yz}\nonumber\\
&&\quad+ S_zI_xA_{zx} + S_zI_yA_{zy} + S_zI_zA_{zz}.
\end{eqnarray}
Writing $x$ and $y$ components of the electron and nuclear spin operators in terms of spin ladder operators, i.e.,
\begin{eqnarray}
S_x=\frac12\left(S_+ + S_-\right),\qquad S_y=\frac{-i}2\left(S_+ - S_-\right),\\
I_x=\frac12\left(I_+ + I_-\right),\qquad I_y=\frac{-i}2\left(I_+ - I_-\right),
\end{eqnarray}
we can write
\begin{eqnarray}
4\textbf{S}\cdot\textbf{A}\cdot\textbf{I}&&= (S_+ + S_-)(I_+ + I_-)A_{xx} - i(S_+ + S_-)(I_+ - I_-)A_{xy}\nonumber\\
&&\quad- i(S_+ - S_-)(I_+ + I_-)A_{yx} - (S_+ - S_-)(I_+ - I_-)A_{yy}\nonumber\\
&&\quad+ 2(S_+ + S_-)I_z A_{xz} -2 i(S_+ - S_-)I_z A_{yz}\nonumber\\
&&\quad+ 2 S_z(I_+ + I_-)A_{zx} -2i S_z(I_+ - I_-) A_{zy} + 4S_zI_zA_{zz}.
\end{eqnarray}
After some straightforward calculations and taking into account that $A_{ij}=A_{ji}$, for $i\ne j$, with $i$ and $j$ being $x,y,z$, we obtain
\begin{eqnarray}
4\textbf{S}\cdot\textbf{A}\cdot\textbf{I}&&=4S_zI_zA_{zz}\nonumber\\
&&\quad+ (A_{xx}+A_{yy})(S_+I_-+S_-I_+)\nonumber\\
&&\quad+ S_+I_+(A_{xx}-A_{yy}-2iA_{xy})\nonumber\\
&&\quad+ S_-I_-(A_{xx}-A_{yy}+2iA_{xy})\nonumber\\
&&\quad+2 (S_+I_z+S_zI_+)(A_{xz}-iA_{yz})\nonumber\\
&&\quad+2 (S_-I_z+S_zI_-)(A_{zx}+iA_{zy}).
\end{eqnarray}

The components of the hyperfine matrix are given as \cite{Nizovtsev_sup}
\begin{eqnarray}
A_{xx}&=&A_c-A_d\left(1-3\sin^2{\theta}\cos^2{\varphi}\right), \label{eq_Axx}\\
A_{yy}&=&A_c-A_d\left(1-3\sin^2{\theta}\sin^2{\varphi}\right),\\
A_{zz}&=&A_c-A_d\left(1-3\cos^2{\theta}\right),\\
A_{xy}&=&A_{yx}=3A_d\sin^2{\theta}\cos{\varphi}\sin{\varphi},\\
A_{xz}&=&A_{zx}=3A_d\sin{\theta}\cos{\theta}\cos{\varphi},\\
A_{yz}&=&A_{zy}=3A_d\sin{\theta}\cos{\theta}\sin{\varphi}, \label{eq_Ayz}
\end{eqnarray}
where $\theta$ and $\varphi$ are the polar and azimuthal angles that determine the nuclear spin, shown in Fig.~1 of the main text. Here, to keep the notation simple we have removed the index $H$ of the angles.
Therefore we can write
\begin{eqnarray}
A_{xx}-A_{yy}\pm 2iA_{xy}=3A_d\sin^2{\theta}e^{\pm i2\varphi},\\
A_{xz}\pm iA_{zy}=3A_d\cos{\theta}\sin{\theta}e^{\pm i\varphi}.
\end{eqnarray}
The hyperfine term can then be written as
\begin{eqnarray}
4\textbf{S}\cdot\textbf{A}\cdot\textbf{I}&&=4S_z I_z A_{zz}\nonumber\\
&&\quad+(2A_c-A_d(2-3\sin^2{\theta}))(S_+I_-+S_-I_+)\nonumber\\
&&\quad+3A_d\sin^2{\theta}(S_+I_+e^{-2i\varphi}+S_-I_-e^{+2i\varphi})\nonumber\\
&&\quad+6A_d\cos{\theta}\sin{\theta}\left((S_+I_z+S_zI_+)e^{-i\varphi} + (S_-I_z+S_zI_-)e^{+i\varphi}\right).
\end{eqnarray}

Defining 
\begin{eqnarray}
\label{eq:CalCoefficients}
&&A_\perp=2A_c+A_d\left(1-3\cos^2{\theta}\right),\\
&&A_{ani}=3A_d\cos{\theta}\sin{\theta},\\
&&A'_\perp=3A_d\sin^2{\theta},
\end{eqnarray}
we obtain
\begin{eqnarray}
\textbf{S}\cdot\textbf{A}\cdot\textbf{I}&&=S_zI_zA_{zz}\nonumber\\
&&\quad+\frac{A_\perp}4(S_+I_-+S_-I_+)\nonumber\\
&&\quad+\frac{A'_\perp}4(S_+I_+e^{-2i\varphi}+S_-I_- e^{+2i\varphi})\nonumber\\
&&\quad+\frac{A_{ani}}2\left((S_+I_z+S_zI_+)e^{-i\varphi} + (S_-I_z+S_zI_-)e^{+i\varphi}\right).
\end{eqnarray}

\section{superoperator master equation}
To find out the evolution of the density matrix of the combined system of an electron and a nuclear spin we use the Markovian master equation \cite{Preskill_sup}
\begin{equation}
\frac{d\rho}{dt}=-i[H,\rho]+\sum^{}_{k}{\mathcal{D}(L_k)\rho},
\end{equation}
in which
\begin{equation}
\mathcal{D}(L_k)\rho=L_k\rho L^{\dagger}_k-\frac{1}{2}\left(\rho L^{\dagger}_k L_k + L^{\dagger}_k L_k \rho\right),
\end{equation}
The first term in the master equation is the unitary evolution of the density matrix due to the total Hamiltonian of the system, while the second term is the evolution due to the interaction with the environment and is called Lindbladian. The operators $L_k$ are known as the jump operators and cause transition between states of the system due to the interaction with the environment. 

In our simulations we included the transitions between the states, due to the optical excitation and spontaneous decay, through the Lindbladian part. For instance, the jump operator that represents the transition from the electron spin state $|i\rangle$ to $|j\rangle$ is given by $\sqrt{\alpha_{ij}}|i\rangle\langle j|\otimes \mathbf{1}$. Here, $\alpha_{ij}$ is the transition rate from the electron spin state $|i\rangle$ to $|j\rangle$ and $\mathbf{1}$ is the nuclear spin identity matrix.  The effect of the inhomogeneous decoherence time, $T^*_2$, can be explained by a jump operator of the form $\frac{1}{\sqrt{2 T^*_2}}S_z \otimes \mathbf{1}$. 

In our simulations we used the vectorized form of the master equation given as \cite{Havel_sup}
\begin{equation}
\frac{d \hat\rho(t)}{dt}= \mathcal{L} \hat\rho(t),
\end{equation}
in which
\begin{equation}
\mathcal{L}=i(\bar H\otimes I-I \otimes H)+\sum^{}_{k}{\left(\bar L_k\otimes L_k-\frac{1}{2}I\otimes L^{\dagger}_k L_k-\frac{1}{2} \bar L^{\dagger}_k \bar L_k\otimes I\right)}.
\end{equation}
Here, $I$ is the $N\times N$ identity matrix where $N=14$ is the dimension of the system of the NV electron and a spin 1/2 nuclear spin. The complex conjugate is shown with over bar and adjoint with dagger. The vector form of the density matrix, $\hat \rho$, is obtained by stacking the columns of the density matrix from left to right on top of each other. The time evolved density matrix is therefore obtained by
\begin{equation}
\hat \rho (t)= e^{\mathcal{L}t}\hat\rho(0).
\end{equation}

\section{Rabi frequency and optimal microwave time}
In the methods based on the precession of nuclear spin while the electron spin is in $m_s=0$ and while in $m_s=1$, the electronic spin is under a microwave excitation while the nuclear spin is under an external magnetic field ($m_s=0$ method) or the hyperfine interaction ($m_s=1$ method). In order to find an analytical formula for the optimal Rabi frequency and the optimal time for the microwave field, we approximate the Hamiltonian of the electron-nuclear spin with a system of three levels. We consider the three levels which are close to each other energetically. In other words, for the $m_s=0$ method we only consider $|0,\downarrow\rangle$, $|0,\uparrow\rangle$, and $|1,\downarrow\rangle$ and for the $m_s=1$ method $|1,\downarrow\rangle$, $|1,\uparrow\rangle$, and $|0,\uparrow\rangle$ states.

With this approximation, in the microwave rotating frame, the Hamiltonian can be written as
\begin{eqnarray}
H=	\left(
\begin{array}{ccc}
0      & \Delta &\Omega\\
\Delta & 0      & 0\\
\Omega & 0      & 0\\
\end{array}\right)\;.
\label{eq:H3}
\end{eqnarray}
Here, $\Delta$ is the nuclear coupling and $\Omega$ is the electronic coupling. We have also approximated all the diagonal terms with a same value which we have set to zero. This can be achieved through the proper choice of the magnetic field and microwave power. The eigenvalues of this Hamiltonian are
\begin{eqnarray}
\lambda_0&=&0,\\
\lambda_\pm&=&\pm\sqrt{\Delta^2+\Omega^2}=\pm\lambda,
\label{eq:evalued}
\end{eqnarray}
and its eigenvectors are
\begin{eqnarray}
v_0&=&\frac1{\sqrt{\Delta^2+\Omega^2}}\left[\begin{array}{ccc}
0,&\Omega,&-\Delta\end{array}\right],\\
v_\pm&=&\frac1{\sqrt{2(\Delta^2+\Omega^2)}}\left[\begin{array}{ccc}
\sqrt{\Delta^2+\Omega^2},&\pm\Delta,&\pm\Omega\end{array}\right].
\label{eq:eenergy}
\end{eqnarray}
The diagonal bases in terms of the eigenvectors can be written as
\begin{eqnarray}
\left[\begin{array}{ccc}1,&0,&0\end{array}\right]
&=&\frac{v_++v_-}{\sqrt{2}},\\
\left[\begin{array}{ccc}0,&1,&0\end{array}\right]
&=&\frac1{\sqrt{\Delta^2+\Omega^2}}\left(\Delta\frac{v_+-v_-}{\sqrt{2}}+\Omega v_0\right),\\
\left[\begin{array}{ccc}0,&0,&1\end{array}\right]
&=&\frac1{\sqrt{\Delta^2+\Omega^2}}\left(\Omega\frac{v_+-v_-}{\sqrt{2}}-\Delta v_0\right).\label{eq:bdiagonal}
\end{eqnarray}

We can write the initial state for the $m_s=0$ method as $\varphi_{110}(t=0)=[1,1,0]/\sqrt{2}$, and for the $m_s=1$ method by $\varphi_{001}(t=0)=[0,0,1]$. The time evolved states are then given by 
\begin{eqnarray}\label{eq:bdiagonal}
&&\varphi_{110}(t)=\left[
\cos(\lambda t)-i\frac{\Delta\sin(\lambda t)}{\sqrt{\Delta^2+\Omega^2}},\,\,\,\,
\frac{\Delta^2\cos(\lambda t)+\Omega^2}{\Delta^2+\Omega^2}-i\frac{\Delta\sin(\lambda t)}{\sqrt{\Delta^2+\Omega^2}},\,\,\,\,
\frac{\Delta\Omega(\cos(\lambda t)-1)}{\Delta^2+\Omega^2}-i\frac{\Omega\sin(\lambda t)}{\sqrt{\Delta^2+\Omega^2}}\right],\\
&&\qquad\qquad\qquad\qquad\qquad\varphi_{001}(t)=
\left[
-i\frac{\Omega\sin(\lambda t)}{\sqrt{\Delta^2+\Omega^2}},\,\,\,\,
\frac{\Delta\Omega(\cos(\lambda t)-1)}{\Delta^2+\Omega^2},\,\,\,\,
\frac{\Omega^2\cos(\lambda t)+\Delta^2}{\Delta^2+\Omega^2}
\right].
\end{eqnarray}

We note that the highest nuclear polarization is achieved for $\Omega=\pm\Delta$ at times
\begin{eqnarray}
t_n=(2n+1)\frac{\pi}{\sqrt{\Delta^2+\Omega^2}},\qquad n=0,1,2,3,....
\end{eqnarray}
Note that, these conditions does not result in electronic spin polarization. However, as is shown in the main text we achieve electronic polarization at the end of the sequence for each method (See Figs.~4 and 5 of the main text). This is due to the optical excitation at the end of the sequence which allows the electronic spin to mainly polarize to $m_s=0$ state while preserving the nuclear polarization.

\section{ESLAC}
In this section we give some details about the polarization based on the ESLAC method.
\subsection{$A_{ani}$, $A'_{\perp}$ components of the hyperfine interaction}
\begin{figure}[t!]
\centering
    \includegraphics[width=.55\textwidth]{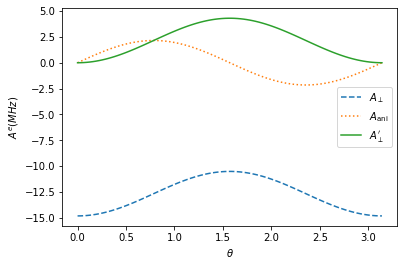}
    \caption{Components of the excited state hyperfine matrix versus angle $\theta$. The hyperfine matrix is constructed by calculating the contact term $A_c$, and dipole coupling $A_d$ for a $^{13}$C nuclear spin in family C setting $\varphi=0$.}
     \label{fig_Amatrix}
\end{figure}

In the ESLAC method, the polarization of the nuclear spin is achieved due to the $A_{\perp}$ component of the hyperfine interaction (see Fig.~3(a) of the main text). However, the other components of the hyperfine interaction, i.e., $A_{ani}$ and $A'_{\perp}$ contribute to the depolarization of the nuclear spin. The anisotropic component causes precession between $|-1,\downarrow\rangle$ and $|-1,\uparrow\rangle$, between $|-1,\downarrow\rangle$ and $|0,\downarrow\rangle$, and between $|-1,\uparrow\rangle$ and $|0,\uparrow\rangle$. On the other hand, $A'_{\perp}$ component causes precession between $|-1,\downarrow\rangle$ and $|0,\uparrow\rangle$. It can be seen in Fig.~6(a) of the main text that for the ESLAC method, the nuclear polarization decreases at angles close to $\theta=\pi/2$. For this range of $\theta$, the $A^\prime_{\perp}$ component of the hyperfine interaction is larger. Figure \ref{fig_Amatrix} shows the components of the excited state hyperfine matrix for a nuclear spin with contact and dipole couplings, $A_c$, and $A_d$, respectively, taken from $^{13}$C family C. Setting the azimuthal angle to $\varphi=0$, the contact term $A_c$ and the dipole coupling $A_d$ can be calculated as
\begin{equation}
A_c=(A_{xx}+A_{yy}+A_{zz})/3, \qquad A_d=A_{xx}+A_{zz}-2A_c.
\end{equation}
The hyperfine matrix is then constructed for various angles $\theta$ using Eqs.~\eqref{eq_Axx} to \eqref{eq_Ayz}.



\subsection{Comparison with the experimental data for $^{15}$N}
\begin{figure}[h!]
\centering
    \includegraphics[width=.95\textwidth]{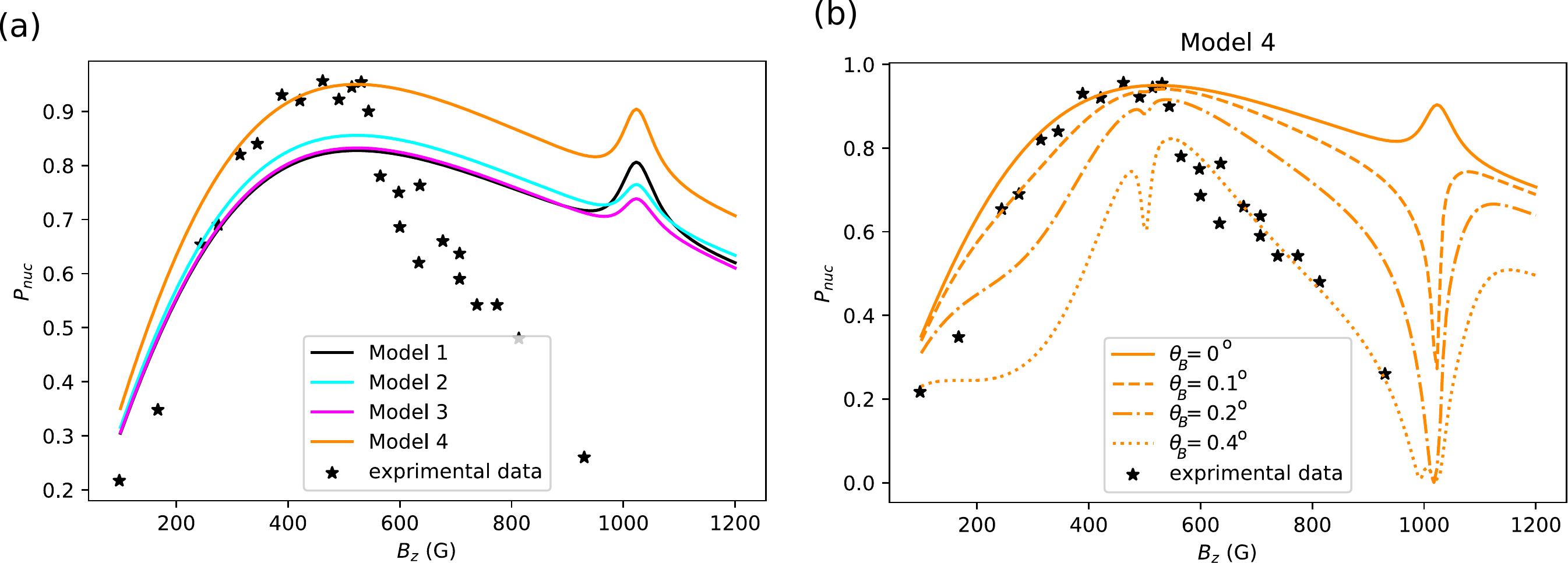}
    \caption{$^{15}$N nuclear polarization for ESLAC method versus $B_z$ for all the 4 transition rate models given in the main text, Table I (a) and for model 4 for a range of angles of the magnetic field $\theta_B$ (b). The experimental data (black stars) are taken from Ref.~\cite{Vincent_sup}.}
     \label{fig_ESLAC}
\end{figure}

We now compare our numerically simulated polarization for the $^{15}$N using the ESLAC method with the experimental data of  Ref.~\cite{Vincent_sup}. Figure \ref{fig_ESLAC} shows the nuclear polarization versus magnetic field at steady state, i.e., for laser excitation time of 1ms. 
Figure \ref{fig_ESLAC}(a) shows models 1 to 4. Model 4 gives the highest nuclear polarization, resulting in $\approx 95\%$ at ESLAC, $B \approx 520$ G, very close to experimental data.
 The discrepancy between the experimental data and our numerical simulations for other values of $B_z$ may be due to the misalignment of the magnetic field as is shown in Fig.~\ref{fig_ESLAC}(b).

\subsection{Comparison with the experimental data for $^{13}$C families}
\begin{figure}[h!]
\centering
    \includegraphics[width=.95\textwidth]{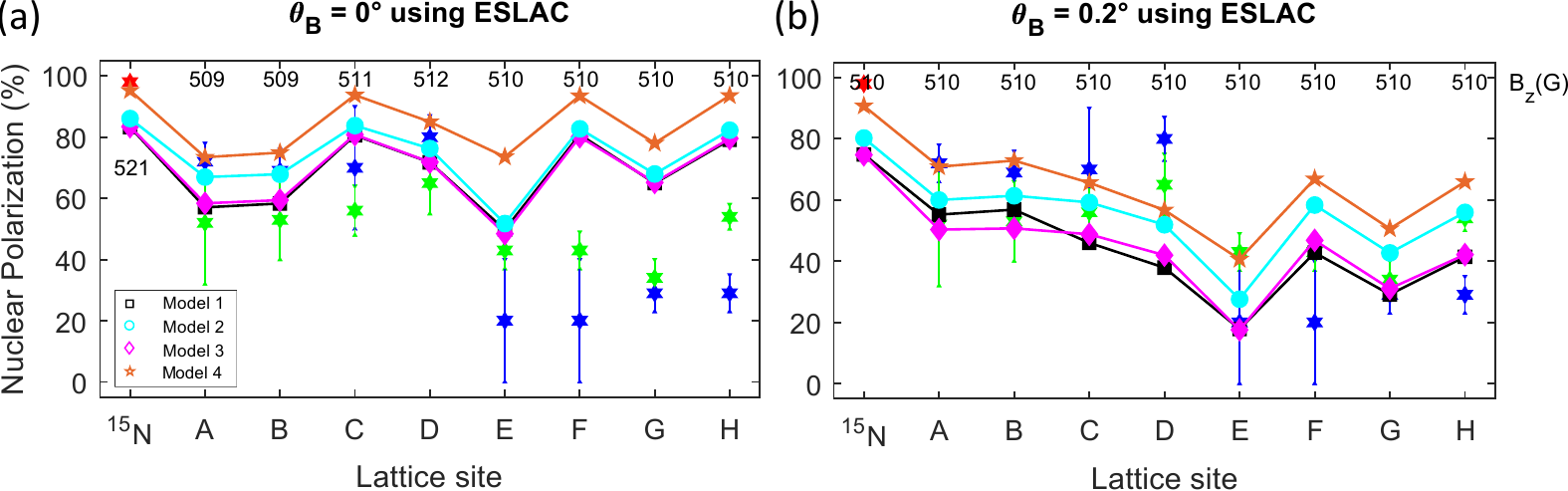}
    \caption{Nuclear spin polarization for $^{15}$N and $^{13}$C families (A to H) for a magnetic field aligned with the NV axis, $z$ axis (a) and for a magnetic field misaligned with the angle $\theta_B=0.2$ degrees. The experimental data for $^{15}$N (red star) is taken from Ref.~\cite{Vincent_sup}, and for $^{13}$ C families are taken from Refs.~\cite{Smeltzer_sup} (blue stars) and \cite{Dreau_sup} (green stars).}
     \label{fig_ESLAC_families}
\end{figure}

Here, we compare our results for the nuclear spin polarization achieved with the ESLAC method for $^{15}$N and families A to H of $^{13}$C with the experimental data of Refs.~\cite{Vincent_sup, Dreau_sup, Smeltzer_sup}. As can be seen in Fig.~\ref{fig_ESLAC_families}(a), for $^{15}$N and for families A to D our results are in good agreement with the experimental data. However, our simulated data result in a higher nuclear polarization for families E to H. This discrepancy could be due to the misalignment of the magnetic field in the experiments. In Fig.~\ref{fig_ESLAC_families}(b) we have shown that even for a small misaligned angle of $0.2$ degrees for the magnetic field, the nuclear polarization reduces significantly for families C to H.

We note that we have taken the hyperfine matrix for such nuclear spin families from Ref.~\cite{Ivady_sup}. The hyperfine matrix given in that reference is given in the principal axis system of the nuclear spin. The angle $\theta$ between the $z$ axis of the nuclear spin, which is in the direction of the eigenvector corresponding to $A_{zz}$, and the $z$ axis of the NV center is also given in that work. Setting the azimuthal angle, $\varphi_H$ in the main text, to zero, a rotation around the $y$ axis with angle $\theta$ gives the hyperfine matrix in the principal axis system of the NV \cite{Nizovtsev10_sup}. Note that, this angle $\theta$ is distinct from angle $\theta_H$ of the main text.

\section{Second order correction}
In this section we show how nonsecular terms of the Hamiltonian can be taken into account with second order perturbation theory. The Hamiltonian of the NV electron spin and a nuclear spin can be written as 
\begin{equation}\label{eq_H}
H = D S_z^2 + \gamma_{el} \mathbf{B}\cdot \mathbf{S} + \gamma_n \mathbf{B}\cdot \mathbf{I} + \mathbf{S} \cdot \mathbf{A} \cdot \mathbf{I},
\end{equation}
where $D$ is the zero field splitting, $\gamma_{el}$ and $\gamma_n$ are the gyromagnetic ratios of the electron and nuclear spin, $\mathbf{A}$ is the hyperfine matrix and $\mathbf{S}$ and $\mathbf{I}$ are the electron and nuclear spin operators. 

The secular terms of the above Hamiltonian can be written as 
\begin{equation}\label{eq_H0}
H_{0} = D S_z^2 + \gamma_{el} B_z S_z + \gamma_n B_z I_z + \sum^{}_{j}{S_z A_{zj} I_j},
\end{equation}
while the nonsecular terms are 
\begin{equation}
V= \gamma_{el} B_x S_x+ \gamma_{el} B_y S_y + \sum^{}_{j}{(S_x A_{xj} I_j+S_y A_{yj} I_j)}.
\end{equation}
We take the nonsecular terms as a perturbation. Using the spin ladder operators $S_{\pm}$ and defining $B_{\pm}=B_x\pm i B_y$ and $A_{j\pm}=A_{jx}\pm i A_{jy}$, the nonsecular terms can be written as
\begin{equation}\label{eq_V}
V=\frac{\gamma_{el}}{2}(S_{+}B_{-}+S_{-}B_{+})+\frac{1}{2}\sum^{}_{j=x,y,z}{(S_{-} A_{j+}+S_{+} A_{j-})I_j}.
\end{equation}

From the second order perturbation theory we have \cite{Sakurai_sup}
\begin{equation}
E^{(2)}_n=\langle n^{(0)}|V \frac{\phi_n}{E^{(0)}_n-H_0} V |n^{(0)}\rangle,
\end{equation}
where $|n^{(0)}\rangle$ are the eigenstates of the unperturbed Hamiltonia, $H_0$, i.e., the eigenstates of $S_z$ spin operator $|m_s\rangle$, and  
\begin{equation}
\phi_n=1-|n^{(0)}\rangle\langle n^{(0)}|=\sum_{k\ne n}{|k^{(0)}\rangle\langle k^{(0)}|}.
\end{equation}

Therefore, for $m_s=0$ we obtain
\begin{equation}
E^{(2)}_{m_s = 0}=\frac{\langle m_s=0|V|m_s=1\rangle\langle m_s=1|V|m_s=0\rangle}{E^{(0)}_0-E^{(0)}_1}+\frac{\langle m_s=0|V|m_s=-1\rangle\langle m_s=-1|V|m_s=0\rangle}{E^{(0)}_0-E^{(0)}_{-1}}.
\end{equation}
where $E^{(0)}_0-E^{(0)}_{\pm 1}=-D\mp\gamma_{el} B_z$ and
\begin{eqnarray}
&&\langle0|V|\pm 1\rangle=\frac{1}{\sqrt{2}}\gamma_{el} B_{\pm}+\frac{1}{\sqrt{2}}\sum_{j}{A_{j\pm}I_j},\\
&&\qquad\langle\pm1|V|0\rangle=\langle0|V|\pm 1\rangle^{\star}.
\end{eqnarray}

After some calculations we obtain 
\begin{equation}\label{eq_E20}
E^{(2)}_{m_s=0}=\frac{-2 D \hat M}{2(D^2-\gamma^2_{el} B^2_z)}+\frac{2\gamma_{el} B_z \hat N}{2(D^2-\gamma^2_{el} B^2_z)},
\end{equation}
where we have defined 
\begin{eqnarray}
&&\hat M= 2\gamma_{el}\left[\left(A_{xx}B_x+A_{yx}B_y\right) I_x \right.+ \left(A_{xy}B_x+A_{yy}B_y\right)I_y+\left. \left(A_{xz}B_x+A_{yz}B_y\right) I_z\right]+ \gamma^2_{el} B^2_{\perp}\mathbf{1}+(\vec A_{+}\cdot \vec A_{-})\mathbf{1}, \\
&&\qquad\qquad\qquad\qquad\qquad\qquad\qquad\qquad \hat N= i (\vec A_+\times \vec A_-)\cdot \vec I.
\end{eqnarray}
Here, $\mathbf{1}$ is the $2\times 2$ identity matrix, and $\vec I$ is a vector whose elements are the Pauli matrices $\vec I=(I_x, I_y, I_z)$.
In the calculations we have used the following identity for Pauli matrices
\begin{equation}
(\vec  A_+ \cdot \vec I)(\vec A_- \cdot \vec I)= (\vec A_+\cdot \vec A_-) \mathbf{1}+ i  (\vec A_+ \times \vec A_-) \cdot \vec I.
\end{equation}

Similarly, for $m_s=\pm 1$ we have
\begin{equation}
E^{(2)}_{m_s = \pm 1}=\frac{\langle \pm 1|V|0\rangle\langle 0|V|\pm 1\rangle}{E^{(0)}_{\pm 1}-E^{(0)}_0},
\end{equation}
which gives 
\begin{equation}\label{eq_E2pm}
E^{(2)}_{m_s=\pm1}=\frac{\hat M\mp \hat N}{2(D\pm \gamma_{el} B_z)}.
\end{equation}
To add the energies given in Eqs.~\eqref{eq_E20} and \eqref{eq_E2pm} to the eigenenergies of the unperturbed Hamiltonian we add the following term to the unperturbed Hamiltonian, Eq.~\eqref{eq_H0},
\begin{eqnarray}\label{Hsoc}
H_{\rm soc}&&=\frac{(3S^2_z-2)D+S_z\gamma_e B_z}{2(D^2-\gamma^2_{el} B^2_z)}\hat M+\frac{(2-S^2_z)\gamma_{el} B_z-S_z D}{2(D^2-\gamma^2_{el} B^2_z)}\hat N.
\end{eqnarray}
The above equation is the second order correction term given in Eq.~(8) of the main text.